\documentclass[preprint]{revtex4}
\usepackage{graphicx}
\usepackage{dcolumn}
\usepackage{amsmath,amsfonts}

\begin{document}

\title{Noise control by sonic crystal barriers made of recycled materials}
\author{Jos\'e S\'anchez-Dehesa}
\email{jsdehesa@upvnet.upv.es}
\author{Victor M. Garcia-Chocano, Daniel Torrent,Francisco Cervera, Suitberto Cabrera}
\affiliation{Grupo de Fen\'omenos Ondulatorios, Departamento de Ingenieria Electr\'onica, Universidad Polit\'ecnica de Valencia, Camino de vera s/n, E-46022 Valencia, Spain}
\author{Francisco Simon}
\affiliation{Departamento de Ac\'ustica Ambiental, Instituto de Ac\'ustica-CSIC, C/ Serrano, 144, E-28006 Madrid, Spain}
\date{\today}

\begin{abstract}
A systematic study of noise barriers based on sonic crystals made of cylinders that use recycled materials like absorbing component is here reported.
The barriers consist of only three rows of perforated metal shells filled with rubber crumb. 
Measurements of reflectance and transmittance by these barriers are reported.
Their attenuation properties result from a combination of sound absorption by the rubber crumb and reflection by the periodic distribution of scatterers. It is concluded that porous cylinders can be used as building blocks whose physical parameters can be optimized in order to design efficient barriers adapted to different noisy environments.
\end{abstract}
\pacs{43.50.Gf, 43.20.Fn}
\maketitle
\section{\label{sec:intro}Introduction}

Sonic Crystals (SCs) are structures made of sound scatterers periodically arranged in a lattice\cite{Dow92}. 
SCs forbid sound propagation for some frequency bands, named bandgaps, in a manner similar as semiconductors forbid the transmission of electronic waves of some energy bands\cite{AscBook}. The physical mechanism behind the formation of bandgaps is the same for both scalar waves; that is, the destructive interference between waves reflected by two consecutive planes of sound scatterers (for acoustic waves) or atomic planes (for electronic waves). 
Transmittance and reflectance measurements for sound waves impinging arrays of solid cylinders in air demonstrated that, at bandgap frequencies, a low transmittance and a high reflectance are simultaneously observed due to Bragg scattering\cite{San98,RubJLT99,San01}. The existence of complete bandgaps as well as the presence of deaf modes in the acoustic bands were demonstrated in the late eighties\cite{San98,RubJLT99}. 
Also, it has been shown that disordering in the SC lattice produces bandgap enlargement\cite{Cab99}. 

Numerical algorithms have been developed to reproduce the experimental findings. They are based on different theoretical approaches like plane-wave expansion \cite{Kus97}, variational methods \cite{San98,RubJLT99,Cab99}, transfer matrix \cite{Sig96}, multiple scattering \cite{San01,Che01,BikPRE03,SanPRB03}, finite differences in time domain (FFTD)\cite{Sig00} and the finite element method\cite{Lan95}.

Practical devices like, for example, acoustic filters or waveguides based on SCs have been proposed and demonstrated \cite{Miy01,Khe04}. 
Taking advantage of the small acoustic impedance of SC at low frequencies, it is also possible to make convergent lenses \cite{Cer02} and Fabry-Perot interferometers \cite{SanPRB03}.
 An application that generates considerable interest in the last years is the noise control by acoustic barriers based on SC.
  Sound attenuation up to 20dB \cite{San98} and 25dB \cite{San02} were obtained with arrangements of metallic cylinders in air and it has been concluded that SC can compete with mass law-based sound screens with the advantage of less volume and weight. 
  It was also predicted that sound attenuation (in dB) of a 2D SC increases linearly with the number of rows\cite{Gof03}, but diffraction effects associated with the finite height of barriers defines a limit to its potential attenuation. 
  Selective noise reduction has been reported by SC barriers based on three-dimensional scatterers \cite{Bat02}. 
  SC barriers with embedded resonances have been also proposed to attenuate efficiently the low frequency region of the audible spectrum (below 500 Hz)\cite{HoAPL03} 

SC barriers made of metallic or rigid cylinders in air show a strong attenuation of the transmitted sound only at bandgap frequencies. In order to overcome this drawback a porous covering of the cylinders has been proposed as a solution to get a more uniform attenuation spectrum\cite{UmnJASA05}.
Umnova and coworkers reported the insertion loss spectra by barriers made of three rows of cylinders with diameter 0.635 cm and lattice constant 1.5 cm, having a very thin porous covering (0.175 cm thick) in a very wide frequency region (up to 50kHz). 
From our point of view it is of great interest to know what is the physical mechanism, absorption or Bragg reflexion, controlling the insertion loss in acoustic barriers based on SCs. Moreover, it is also important to know how attenuation depends on the thickness of the covering layer and to study this dependence in the low frequency region of the audible spectrum. Both effects are relevant when we consider the design of barriers where the reflected sound needs to be minimized. Thus, for the cases of narrow roads or railway lines, where barriers are placed at both sides, sound reflected by one of the barriers can reach the other with the consequent loss of efficiency.

In this work we report a systematic study, theoretical and experimental, of SC barriers whose attenuation properties are based on a combination of sound absorption by the porous covering and sound reflection by the periodic distribution of scatterers. For this purpose we have fabricated barriers whose building blocks consist of hollow and perforated metallic cylinders filled with rubbed crumb, a porous material that is obtained by recycling used car tires.   
We analyze barriers made of only three rows of cylinders arranged in a square lattice. Moreover, we focus our attention in the low frequency region of the audible spectrum (up to 5kHz) since the barrier quality depends on its performance in this region.

The article is organized as follows. Section II gives a brief review of the multiple scattering method employed in the numerical simulations. Section III describes the acoustical parameters of rubber crumb, the recycled material used as the porous material in building the cylindrical units. The experimental set up and barrier characterization is described in Sec. IV, where a comparison between theory and experiments is also reported together with a discussion of results. Finally, Sec. V gives a summary of the work performed and suggests lines of future research.
\section{\label{MSmodel}Multiple Scattering theory}
Numerical simulations are performed in the framework of multiple scattering theory. 
This theory is based on the exact solution of the single scattering problem and is capable of calculating the scattering of finite and infinite arrays of scatterers. 
A detailed account of the algorithm can be found, for example, in Ref. \onlinecite{SanPRB03} and references therein.

Three different sound scatterers have been tested in the experiments: i) cylindrical rods made of metal, ii) cylinders made of rubber crumb, and iii) cylinders consisting of a shell of rubber crumb and a metal core. 
This section gives a brief account of the expressions for the $t$ matrix of these three types of cylindrical scatterers and how they are obtained. 
Due to the high acoustic impedance of steel compared to air this material can be considered as a rigid body; i.e. with an infinite mass density. Regarding the rubber crumb, it is modeled as a fluid-like medium characterized by three parameters: porosity, complex wave number and complex dynamical mass density. 
Section \ref{seccionA} obtains the $t$ matrix for a porous cylinders and a rigid (metallic) cylinder, respectively. Finally, the $t$ matrix of a porous cylindrical shell with a rigid core is reported in Sec. \ref{seccionB}.

\subsection{\label{seccionA}T matrix of a fluidlike porous cylinder}
Let us consider a fluidlike cylinder of infinite length, radius $R$, mass density $\rho_a$ and sound velocity $c_a$.
It is embedded in a background with acoustic parameters $\rho_b$ and $c_b$.

The general expressions for the pressure field incident on the cylinder $P^0$, scattered by the cylinder $P^{SC}$ and transmitted inside the cylinder $P^{in}$, are:

\begin{eqnarray}
\label{P0}
P^0(r,\theta;\omega)&=&\sum_{q=-\infty}^{+\infty}{A_q^0J_q(k_br)e^{iq\theta}},\quad(r>R)\\
\label{Psc}
P^{SC}(r,\theta;\omega)&=&\sum_{q=-\infty}^{+\infty}{A_qH_q(k_br)e^{iq\theta}},\quad(r>R)\\
\label{Pin}
P^{in}(r,\theta;\omega)&=&\sum_{q=-\infty}^{+\infty}{B_q^0J_q(k_ar)e^{iq\theta}},\quad(r<R)
\end{eqnarray}
where $J_q$ and $H_q$ are q-th order Bessel and Hankel functions, respectively, and $\omega$ is the angular frequency.
Moreover, $k_a=\omega/c_a$ ($k_b=\omega/c_b$) is the wave number inside (outside) the cylinder and $(r, \theta)$ define the polar coordinates of the point where the field is calculated. 

Coefficients $A^0_q$ are defined by the incident field. 
Thus, for a plane wave with amplitude $C$ and wave vector $\vec {k_0}(x, y)=\vec {k_0}(\cos\theta_0,\sin\theta_0)$:
\begin{equation}
\label{A0}
A_q^0=Ci^qe^{-iq\theta_0}
\end{equation}

Coefficients $A_q$ and $B_q$ depend on $A^0_q$ and on the cylinders physical parameters. 
So, to obtain $A_q$ we apply the usual continuity conditions at the cylinders surface\cite{UmnJASA05}:
\begin{subequations}
\begin{eqnarray}
\label{P_bound1}
P^{out}(r=R)&=&P^{in}(r=R),\\
\label{P_bound2}
\frac{1}{\rho_b} \left.\frac{\partial P^{out}}{\partial r}\right|_{r=R}&=&\frac{\Omega}{\rho_a(\omega)} \left.\frac{\partial P^{in}}{\partial r}\right|_{r=R},
\end{eqnarray}
\end{subequations}
where $\rho_a(\omega)$ is the frequency dependent dynamical mass density and $\Omega$ is the cylinder's porosity. The porosity is defined as $\Omega=V_{air}/V_{tot}$, where $V_{air}$ is the pore volume and $V_{tot}$ the total volume of material.

After some easy operations we obtain the so called $t$ matrix of the cylinder, which is defined by

\begin{eqnarray}
\label{Tfluido}
T_q\equiv\frac{A_q}{A_q^0}=-\frac{\rho_qJ_{q}^\prime(k_bR)-J_q(k_bR)}{\rho_qH_{q}^\prime(k_bR)-H_q(k_bR)},
\end{eqnarray}
where 
\begin{equation}
\rho_q=\frac{1}{\Omega}\frac{\rho_ac_a}{\rho_bc_b}\frac{J_q(k_aR)}{J_{q}^\prime(k_aR)}\nonumber
\end{equation}
and $J_q^\prime$ and $H_q^\prime$ are first derivatives of q-th order Bessel and Hankel functions.
This $t$ matrix is used in modeling rubber crumb cylinders. 

For the case of a rigid cylinder ($\rho_a=\infty$), expression \eqref{Tfluido} is reduced to the well know expression
\begin{equation}
\label{Trigido}
T_q=-\frac{J_{q}^\prime(k_bR)}{H_{q}^\prime(k_bR)}
\end{equation}

\subsection{\label{seccionB}T matrix of a porous cylindrical shell with a rigid core}

Let us consider now a cylinder made of a fluid-like porous shell defined by radii $R_a$ and $R_b$ $(R_a<R_b)$ and parameters $k_s$, $\rho_s$, $c_s$. Regarding the core ($r<R_a$), let us start by assuming that is also fluidlike with parameters $k_a$, $\rho_a$, $c_a$ .

The external pressure impinging the cylinder, $P^0$, and the scattered pressure,$P^{SC}$ are given by expressions (\ref{P0}) and (\ref{Psc}), respectively, with $R=R_b$. However, at positions $r<R_b$ we have:

\begin{eqnarray}
\label{Pa}
P^a&=&\sum_{q=-\infty}^{+\infty}{D_qJ_q(k_ar)e^{iq\theta}},\quad {\rm if}\ r<R_a\\
P^s&=&\sum_{q=-\infty}^{+\infty}{B_qJ_q(k_sr)e^{iq\theta}}+\sum_{q=-\infty}^{+\infty}{C_qH_q(k_sr)e^{iq\theta}},\quad {\rm if} \ R_a<r<R_b
\label{Ps}
\end{eqnarray}

By applying the boundary conditions at the two interfaces and also by using the simplifying assumption of a rigid core ($\rho_a\to\infty$), we get:

\begin{equation}
\label{Tcoronafluida}
T_q=-\frac{\rho_wJ_{q}^\prime(k_bR_b)-J_q(k_bR_b)}{\rho_wH_{q}^\prime(k_bR_b)-H_q(k_bR_b)},
\end{equation}
where
\begin{eqnarray}
\label{rhow}
\rho_w&&= \frac{1}{\Omega}\frac{\rho_sc_s}{\rho_bc_b}\frac{J_q(k_sR_b)+T_q^sH_q(k_sR_b)}{J_{q}^\prime(k_sR_b)+T_q^sH_q^\prime(k_sR_b)}\nonumber \\
\label{Tqs}
T_q^s&&= -\frac{J_q^\prime(k_sR_a)}{H_q^\prime(k_sR_a)}\nonumber
\end{eqnarray}

These expressions define the $t$ matrix of a porous cylinder with a rigid core that will be used in the multiple scattering simulations described below.
%
\subsection{\label{multiplescattering}Multiple Scattering}
\begin{figure}
\includegraphics{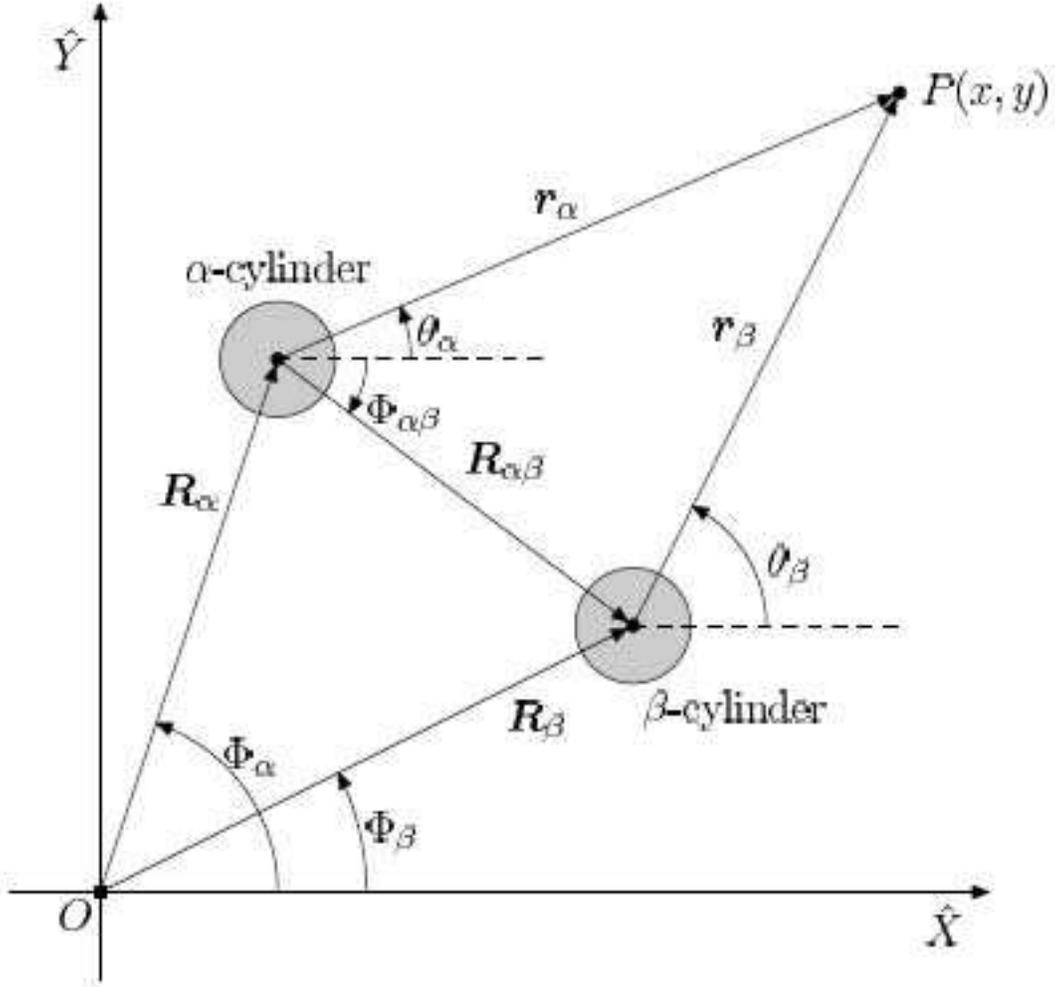}
\caption{\label{CoordinateSystem}Coordinate system and variables used in the multiple scattering theory.}
\end{figure}

When the external field $P^0$ impinges a cluster of N parallel cylinders arbitrarily located, the scattered field $P^{SC}$ is obtained from
\begin{eqnarray}
\label{Pscmult}
P^{SC}(x,y)=\sum_{\alpha=1}^{N}{\sum_{q=-\infty}^{+\infty}{(A_\alpha)_qH_q(k_br_\alpha)e^{iq\theta_\alpha}}},
\end{eqnarray}
where $(r_\alpha, \theta_\alpha)$ are the polar coordinates translated to the center of the $\alpha$th-cylinder  and $(A_\alpha)_q$ the coefficients to be calculated. See Fig. (\ref{CoordinateSystem}) for an account of the variables employed below.

The total field incident on the $\alpha$th-cylinder is
\begin{eqnarray}
\label{Palpha0}
P_\alpha^0=\sum_{s=-\infty}^{+\infty}{(B_\alpha)_sJ_s(k_br_\alpha)e^{is\theta_\alpha}}.
\end{eqnarray}
This takes into account the external field as well as the field scattered by the rest of cylinders arriving at $\alpha$.

Applying Graff's addition theorem to \eqref{Pscmult} and adding the incident field $P^0$ we arrive at

\begin{eqnarray}
P_\alpha^0=\sum_{q=-\infty}^{+\infty}(A_\alpha^0)_qJ_q(k_br_\alpha)e^{iq\theta_\alpha}+
\sum_{\beta\ne\alpha}(A_\beta)_sH_{q-s}(k_br_{\alpha\beta})e^{i(s-q)\theta_{\alpha\beta}}J_q(k_br_\alpha)e^{iq\theta_\alpha}
\label{Palpha0graff}
\end{eqnarray}
where $(A^0_\alpha)_q$ are the same as $A^0_q$ of \eqref{P0} but translated to the $\alpha$th-cylinder. 

For the case of an external plane wave like that in \eqref{A0}
\begin{equation}
\label{A0alpha}
A^0_\alpha=Ci^qe^{-iq\theta_0}e^{i\vec k_0\cdot\vec r_\alpha}
\end{equation}

The $t$ matrix relates $(A_\alpha)_q$ and $(B_\alpha)_s$:
\begin{equation}
\label{AeqTB}
(A_\alpha)_q=\sum_{s=-\infty}^{+\infty}{(T_\alpha)_{qs}(B_\alpha)_s}
\end{equation}
The expression for $(A_\alpha)_q$ is finally obtained from Eqs. \eqref{Palpha0}, \eqref{Palpha0graff}, and \eqref{AeqTB} by truncating the infinite sums at $\pm S_{max}$:
\begin{equation}
\label{Aalpha}
(A_\alpha)_q=\sum_{\beta=1}^{N}{\sum_{r=-S_{max}}^{+S_{max}}{\sum_{s=-S_{max}}^{+S_{max}}{(M_{\alpha\beta}^{-1})_{qr}(T_\alpha)_{rs}(A_\alpha^0)_s}}},
\end{equation}
where
\begin{eqnarray}
\label{M}
(M_{\alpha\beta})_{qs}&=&\delta_{rs}\delta_{\alpha\beta}-(G_{\alpha\beta})_{rs}\\
(G_{\alpha\beta})_{rs}&=&(1-\delta_{\alpha\beta})(T_\alpha)_qH_{q-s}(k_br_{\alpha\beta})e^{i(s-q)\theta_{\alpha\beta}}
\label{G}
\end{eqnarray}
and $\delta$ represents the Kronecker delta. 

Once $(A_\alpha)_q$ is known the total pressure at any point $(x,y)$ can be obtained by adding $P^0(x,y)+P^{SC}(x,y)$. 
Note that this method allows to deal with cylinders of different acoustic parameters at the different positions in the lattice.

%
\section{\label{ParametersRC}Rubber Crumb Parameters}
%
\begin{figure}
\includegraphics{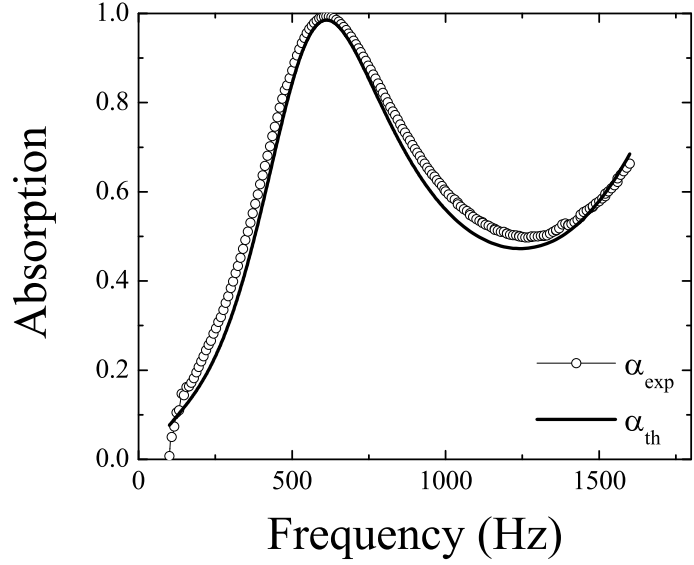}
\caption{\label{fig:alfaGranza} The absorption value of rubber crumb measured in a standing wave tube. The theoretical coefficient has been obtained by fitting $\beta_1$ and $\beta_2$ in Eq. \eqref{alfateo}. The curve $\alpha_{th}$ is obtained with $\beta_1=$1.58 and $\beta_2=$0.7}
\end{figure}

Porous materials are widely used for sound absorption because of their capacity to dissipate the acoustic energy\cite{AttPR82}. 
Granular porous material can be considered as an alternative to the more usual fibrous and foam absorbers. 
In addition, granular materials offer good absorption combined with good mechanical strength unlike fibrous materials\cite{VorAA03}.
Particularly, the rubber crumb is widely used for acoustic conditioning in outdoors noise control\cite{Sim04}. 
The one used here is polydisperse and consists of grains with sizes in the range from 0 to 3mm.  

There are many theoretical and empirical models that have been developed to predict the acoustical properties of granular porous materials. For a review the reader is addressed, for example, to Refs. \onlinecite{AlaBook93,AttAA93,StiJASA92,ChaJASA92,JohJPM87} and references therein.
Here, we employed the so called Biot-Allard theory, which assumes a negligible contribution by the movement of the material skeleton\cite{Ala90}. 
In this model the porous material is described like a dissipative compressible fluid whose acoustical properties are completely characterized by the wave number $k_c(\omega)$ and the characteristic impedance $Z_c(\omega)$, which are functions of $\omega$.
Both are related to the complex dynamical mass density, $\rho(\omega)$,and bulk modulus, $K(\omega)$, through\cite{ChaJASA92}:
\begin{subequations}
\begin{eqnarray}
\label{ks}
k_c(\omega)&=&\omega\sqrt{\frac{\rho(\omega)}{K(\omega)}},\\
\label{Zs}
Z_c(\omega)&=&\frac{1}{\Omega}\sqrt{\rho(\omega) K(\omega)},
\end{eqnarray}
\end{subequations}
where $\Omega$ is the material porosity already introduced in section \ref{seccionA}
%
%
\begin{figure}
\includegraphics{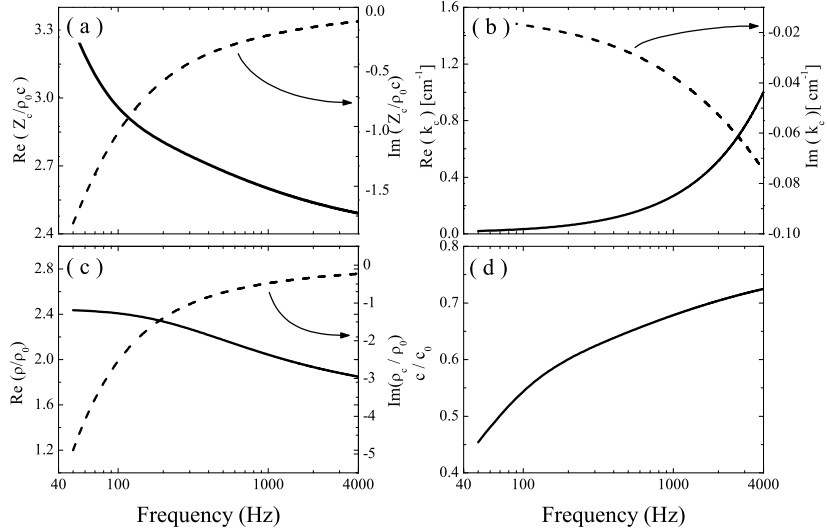}
\caption{\label{fig:KZgranza} Real and imaginary parts of: (a) characteristic impedance $Z_c$; (b) propagating wave vector $k_c$; (c) dynamical mass density $\rho$; and (d) effective sound speed $c$. Note that parameters are determined from density $\rho$ and bulk modulus $K$, which are the two independent parameters in the model (see text).}
\end{figure}

The magnitudes $k_c$ and $Z_c$ are modeled in the present work by the five parameter model of Johnson-Stinson that uses the proposal of Johnson {\it et al.}\cite{JohJPM87} to express the complex dynamical mass density:
\begin{equation}
\label{rhocomplex}
\rho(\omega)=\rho_0\tau\left (1-iA\sqrt{1+i\frac{\beta_1^2}{2A}}\right )\ \ {\rm with} \ A=\frac{\sigma\Omega}{\omega\rho_0\tau},
\end{equation}
where this expression has been obtained by using the viscous characteristic length, $\Lambda$, given in Ref. \onlinecite{ChaJAP91}:
\begin{equation}
\Lambda=\frac{1}{\beta_1}\sqrt{\frac{8\eta\tau}{\sigma\Omega}}
\end{equation}
For the bulk modulus we use the expression obtained in Ref.\onlinecite{AllJASA92} from the initial proposal of Stinson {\it et al.}\cite{StiJASA92}:
\begin{equation}
\label{Kc}
K(\omega)=\frac{\gamma P_0}{\gamma-\frac{\gamma-1}{1-i\frac{\beta_2^2}{N_{pr}}A\sqrt{1+i\frac{N_{pr}}{2A\beta_2^2}}}},
\end{equation}
where $N_{pr}=0.706$ is the fluid Prandtl number and $\gamma=1.4$ is the ratio of specific heats.
This expression has been obtained by using the thermal characteristic length, $\Lambda^\prime$, given in Ref. \onlinecite{ChaJAP91}:
\begin{equation}
\Lambda^\prime=\frac{1}{\beta_2}\sqrt{\frac{8\eta\tau}{\sigma\Omega}}
\end{equation}

The flow resistivity ($\sigma$), the tortuosity ($\tau$) and the porosity ($\Omega$) are parameters that can be experimentally determined by non acoustical methods. Finally, the two parameters $\beta_1$ and $\beta_2$ are related with the attenuation of acoustic energy due to the viscous and thermal phenomena occurring as the sound moves through the material and are the only fitted parameters in our modeling.

The flow resistivity is measured by using an experimental setup based on the European Norm\cite{EN29053}. 
The tortuosity is a parameter characterizing the skeleton of absorbing materials, which is directly related to the porous shape and to the existence of lateral branches. 
In order to its determination, we used the procedure developed by Johnson {\it et al.}\cite{JohPRL82}, which is based on an electric measurement proposed by Brown\cite{BroGEO80}. 
The porosity, $\Omega$, is obtained by a method currently employed in geophysical studies\cite{BroGEO80} because of its simplicity and accuracy. 
The values measured for these parameters are $\sigma=$3318.6 MKS rayls/m, $\tau=$1.54, and $\Omega=$54.1$\%$. 

The parameters $\beta_1$ and $\beta_2$ are obtained by fitting the absorption coefficient measured on a sample of rubber crumb with thickness ${\it e}$ in a standing wave tube. At normal incidence, the theoretical coefficient $\alpha_{th}$ is related to the surface impedance $Z_s$ of the sample by:
\begin{equation}
\label{alfateo}
\alpha_{th}=1-\left|\frac{Z_s-Z_0}{Z_s+Z_0}\right|^2;\ \ \ \ Z_s=-iZ_c{\rm cot}(k_ce),
\end{equation}
where $Z_c$ and $k_c$ are known through the Johnson-Stinson model employed to take into account the physical mechanism of the acoustic wave attenuation within the rubber crumb. 

Figure \ref{fig:alfaGranza} shows the absorption coefficient, $\alpha$, measured for the thickness ${\it e}=$9.5cm. It is compared with the curve $\alpha_{th}$ giving the best fitting. The values obtained for $\beta_1$ and $\beta_2$ are $\beta_1=1.58m^{-1}$ and $\beta_2=0.7m^{-1}$. These values give $\Lambda=142 \mu m$ and $\Lambda^\prime=508\mu m$.


The physical parameters [$Z_c(\omega)$, $k_c(\omega)$, $\rho(\omega)$ and $c(\omega)$] describing our specific rubber crumb are depicted in Fig. \ref{fig:KZgranza}, where the sound speed, $c(\omega)$, through the effective fluid is obtained from:
\begin{equation}
\label{ceff}
c(\omega)=Re\left(\sqrt{\frac{K(\omega)}{\rho(\omega)}}\right)
\end{equation}
The frequency dependence of the different curves in Fig. \ref{fig:KZgranza} follows the standard behavior of granular porous media. 
Note that an increasing sound attenuation is expected for increasing frequencies of the waves propagating through this type of absorbing material.

\begin{figure}
\includegraphics[width=9cm]{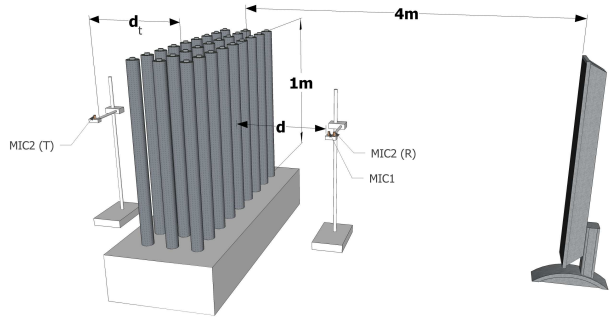}
\caption{\label{fig:setup}Experimental setup. The parameters and dimensions specified here are employed in Section IV. Note that reflectance is obtained by using two microphones put in front of the barrier while transmittance is obtained by using a single microphone [Mic2(T)] put at the rear of the barrier.}
\end{figure}
%
\section{\label{sec:ExpCharac}Experimental characterization}
%
Experiments have been performed in an anechoic chamber of size $8\times 8\times 8 m^3$. 
The samples consist of 3 rows of cylindrical scatterers, each row containing 9 cylinders 1 meter length put in a square configuration with lattice constant $a=$11cm. 

Five different barriers samples have been analyzed and their dimensions are given in Table \ref{tab:Samples}. 
Note that samples 1 and 2 do not contain rubber crumb, they are made of rigid cylinders and results are used here for comparison purposes.
Samples 3, 4 and 5 are made by using 3 different porous units. 
All the units consist of a porous shell with a diameter of $d_b=$8cm. However, their inner cores (steel cylinders) have different diameters: $d_a=$4 cm in sample 3, $d_a=$2cm in sample 4, and $d_a=$0 (no core) in sample 5.


The experimental setup is illustrated in Fig. \ref{fig:setup}. 
As sound excitation we use an UDE AC-150 column loudspeaker separated 4m from the sample in order to have approximately a plane wave front at the surface sample. 
Two B\verb+&+K 4958 microphones are placed at the equatorial plane of samples and aligned with the central cylinder. 
Data of two microphones are acquired with a NI-4551 for further processing in a computer.

Reflectance measurements are performed by placing both microphones between sample and loudspeaker. 
Microphones (Mic1 and Mic2(R) in Fig. \ref{fig:setup}) are put at distances $d$ and $\ell$ near the sample surface, the separation between them being $\Delta d=\ell-d$=1.5cm. 
Note that reflection coefficient can't be determined at frequencies where distance between microphones corresponds to a half-wavelength of sound. 
Thus there is an upper frequency limit that in our case is 11.3 kHz. 
Applying white noise to the sample, the pressure reflection coefficient is calculated as\cite{ChuI80,ChuII80}
\begin{equation}
\label{r_2micros}
r(\omega)=e^{-2ik\ell}\frac{H_{12}e^{-ik\Delta d}-1}{1-H_{12}e^{ik\Delta d}},\nonumber.
\end{equation}
where $H_{12}$ is the transfer function that is experimentally obtained from $S_1$ and $S_2$, the complex spectra (unit of pressure) for the signals measured by Mic1 and Mic2(R), respectively. 
Explicitly, $H_{12}$ is obtained by calculating the ratio $S_{12}/S_{11}$, where  $S_{11}=S_1^*\cdot S_1$ is the autospectrum of signal at Mic1 and $S_{12}==S_1^*\cdot S_2$ is the crossspectrum between signals of both microphones \cite{ISO10534}. 
Therefore, the power reflection coefficient is \cite{ChuI80,ChuII80,ISO10534}
\begin{equation}
\label{R_2micros}
R(\omega)=\left|r(\omega)\right|^2=\left|e^{-2ik\ell}\frac{S_{12}e^{-ik\Delta d}-S_{11}}{S_{11}-S_{12}e^{ik\Delta d}}\right|^2
\end{equation}
This expression represents a one-dimensional approach to the full two-dimensional problem; i.e., it has been obtained under the assumption that the incident, reflected and transmitted waves have plane wavefronts and they all travel along the direction normal to the sample's surfaces\cite{ChuI80,ChuII80,ISO10534}. 
This hypothesis, which is considered valid in our set up since the loudspeaker is far from the sample, has been widely employed by some of us in previous works\cite{San98,RubJLT99,San01,HakAPL05} and we always found good agreement between data and simulations based on it. 
For example, the reader is addressed to Ref.\onlinecite{HakAPL05} where a comparison is reported between data and two numerical simulations; one using a plane wave for the incident beam, the other using a Gaussian wave approach.  

Transmission measurements are made with Mic2 put at the rear surface of the sample (and switching off Mic1); at a distance $d_t$ from the central cylinder. Keeping in mind that the loudspeaker is distant to the sample, the sound power transmission coefficient $T$ can be approximated as the ratio between the autospectrum measured with the SC sample, $S_{22}$ (related with the power of the transmitted field) and that measured without it $S^0_{22}$ (related with the power of the incident field): 
\begin{equation}
\label{T_1micro}
T(\omega)=\frac{S_{22}}{S_{22}^0}
\end{equation}  
The insertion loss (IL), in decibels, is obtained from:
\begin{equation}
\label{insertionloss}
IL(dB)=-10\times {\rm log}_{10}(T).
\end{equation}

Finally, the absorption is obtained by assuming energy conservation: $A(\omega)=1-T-R$. 
At this point it is important to remark that, since we are working with periodic structures, waves propagating along angles $\theta_n$ different to the incident $\theta_0$ are possible because of their scattering by planes of cylinders (Bragg planes) different to the specular. 
These angles are given by\cite{AscBook}:
\begin{equation}
\label{braggwaves}
{\rm sin}(\theta_n)={\rm sin}(\theta_0)+\frac{2\pi n}{ka}, \ \ 
\end{equation}
where $n$ is an integer ($n=0\,\pm 1\,\pm 2\,\ldots$) and $a$ is the lattice constant. Taking $\theta_0=$0 the first diffracted mode appears when $ka=2\pi$, i.e. when $\lambda=a$, in such a manner that $\theta_n$ becomes a real number. 
This condition is known as the diffraction limit and it defines a frequency cut-off such that, for higher frequencies, Eq. \eqref{R_2micros} is not longer valid since reflected and transmitted waves do exist with $k-$wavevectors not collinear with that of the incident wave. In other words, for wavelengths smaller than $a$ some energy is scattered to angles different to the ballistic and will cause the failure of the measurement method due to the non satisfied condition of incident, reflected and transmitted waves travelling along the same direction in the setup. Therefore, no discussion of data will be performed above the diffraction limit. However, let us stress that, for finite structures, Eq. \eqref{R_2micros} can still be used in the frequency regions corresponding to the first bandgap because it appears at frequencies bellow the diffraction limit and where the transmission reduction observed is caused by the destructive interference of waves reflected on the successive rows of cylinders with the incident wave that travels in opposite direction.

Reflectance and transmittance spectra are taken for frequencies up to 4 kHz and for three different values of $d_t$ and $\ell$; 5, 10 and 20cm, respectively. The results within that range of frequencies are consistent and repetitive for the three positions of microphones. 
Here, we only depict spectra correspond to microphones put at positions $d_t$=$\ell$=10cm. 

The data are compared with theoretical spectra obtained by modeling the same experimental configuration. Numerical simulations are performed by assuming an incident plane wave impinging a SC cluster identical to that in the set up. The pressure is also calculated at the microphone positions.
A frequency sweep has been carried out and Eqs. \eqref{R_2micros} and \eqref{T_1micro} are evaluated at each frequency by replacing $S_{11}$ by $p_1$
, the pressure amplitude measured at Mic1, and $S_{21}$ by $p_2$, the pressure amplitude at Mic2. 
The main difference is that multiple scattering calculations are performed under the approach of infinite long cylinders. This assumption is justified because the data are taken very close to the sample surface and, therefore, no diffraction from the borders of cylinders is expected. In other words, the cylinders effectively behave as infinitely long.
%
%
\begin{figure}
\includegraphics{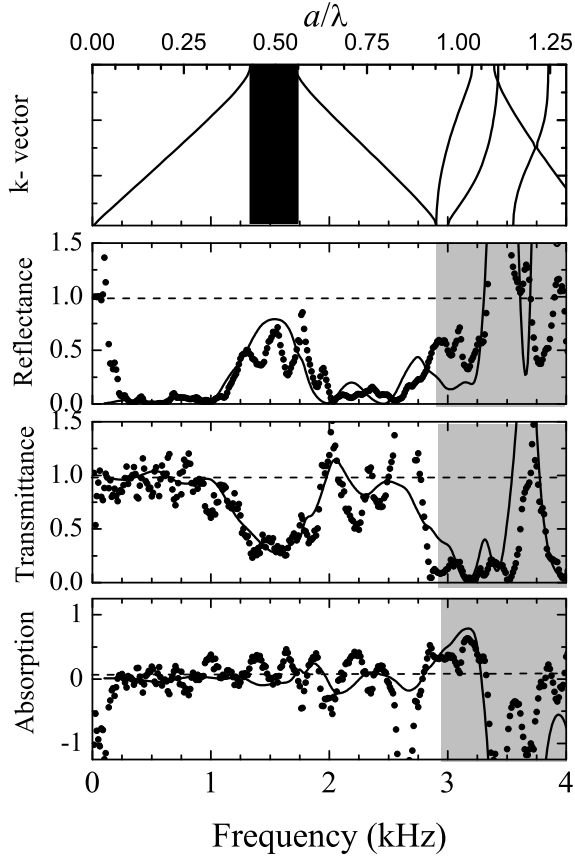}
\caption{\label{GraficaNG10}Reflectance, transmittance and absorption spectra (black dots) for a noise barrier made of three rows of metal cylinders 4 cm diameter (sample 1 in table I). The simulations (solid lines) are performed by using the multiple scattering theory described in Sec. II. The horizontal dashed line is a guide for the eye. The shadowed regions define the frequencies above the diffraction limit. The black stripe in the acoustic band structure (top panel) defines the bandgap of the underlaying square lattice.}
\end{figure}
%
%
\begin{figure}
\includegraphics{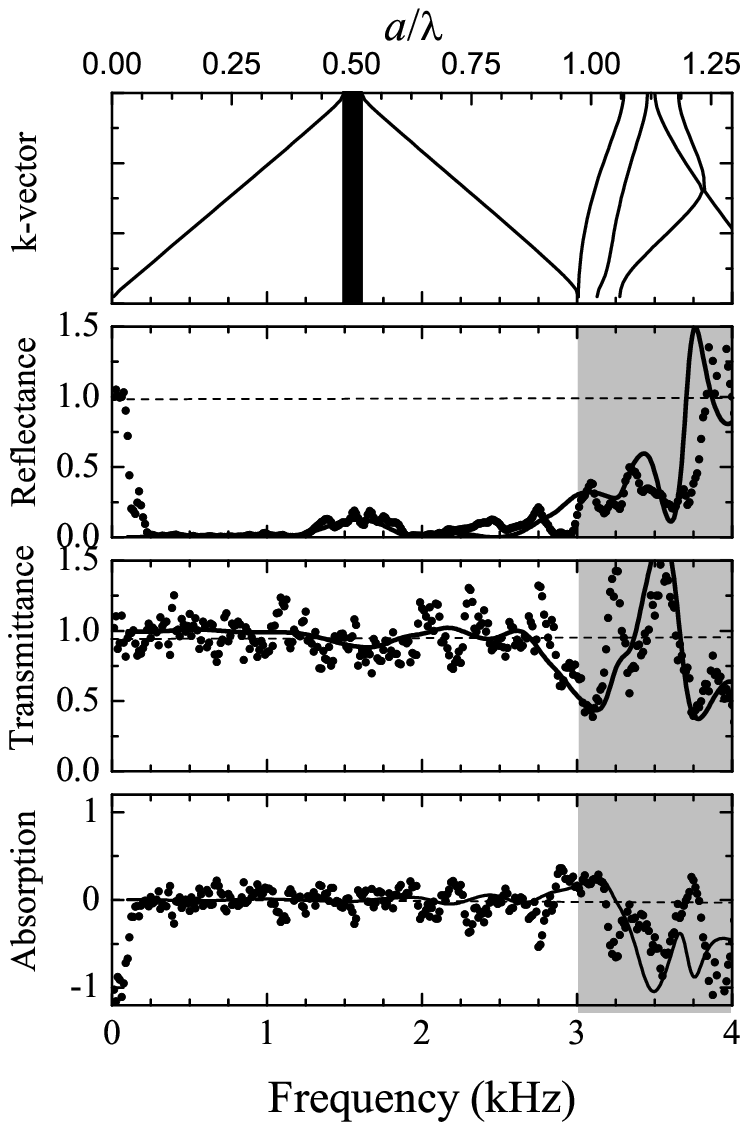}
\caption{\label{GraficaNF5}Reflectance, transmittance and absorption spectra (black dots) for a noise barrier made of three rows of metal cylinders 2 cm diameter (sample 2 in table I). The simulations (solid lines) are performed by using the multiple scattering theory described in Sec. II. The horizontal dashed line is a guide for the eye. The shadowed regions define the frequencies above the diffraction limit. The black stripe in the acoustic band structure (top panel) defines the band gap of the underlaying square lattice.}
\end{figure}
%
\subsection{Barriers made of rigid cylinders}
%
Figures \ref{GraficaNG10} and \ref{GraficaNF5} show the results for SC barriers made of only rigid cylinders with diameters 4cm and 2cm, respectively; i.e., samples 1 and 2 in Table I. 
For the sake of comparison, the band structures along the $k-$direction normal to the sample surface are also shown. 
The acoustic bands have been obtained by the procedure described in Section II of Ref. [10]. 
In brief, we have solved the secular equation obtained by taking into account that, in a periodic system, the Bloch theorem applies to coefficients $(A_\alpha)_q$ given by Eq. \eqref{AeqTB}. 
It is observed that, at the frequency region where the band gap is predicted (black stripes), a peak appears in the reflectance spectrum and a deep simultaneously appears in the transmittance spectrum. 
These features are barely seen in sample 2, which is made of very thin cylinders and, therefore, has a small filling fraction, $ff$. This parameter is defined as the ratio between the cylinder volume and the total volume of the SC unit cell (see Table I). Let us stress that better defined peaks and deeps can be obtained if larger number of layer were employed in building the barrier\cite{San98}. Note that data (black dots) are well reproduced by numerical simulations based on multiple scattering theory  (continuous lines). 

It is noticeable in Figs. \ref{GraficaNG10} and \ref{GraficaNF5} that values larger than unity appear in the reflectance or transmittance spectra for frequencies above 3kHz. These values are unphysical and are obtained by the fact that expressions employed in their calculations lost their validity since they were obtained under the approach of a plane wavefront propagating along the direction normal to the sample surface\cite{ChuI80,ChuII80}. 
 According to Eq. \eqref{braggwaves} the excitation of Bragg waves with $n\ne0$ approximately starts around $\lambda=a$. An exact determination of this cutoff is obtained from the acoustic band structure (top panels in Figs. \ref{GraficaNG10} and \ref{GraficaNF5}), which gives about 2.8kHz for sample 1 and 3kHz for sample 2. These cutoffs are confirmed by the absorption spectra, which are flat (with value zero) up to such frequencies. Since microphones only measure ballistic transport (zero-order Bragg waves) and there is no mechanism for sound absorption in these samples, the non-zero values above such frequencies are mainly due to diffraction effects.  
\par
%
\subsection{Barriers made of rubber crumb cylinders}
%
%
\begin{figure}
\includegraphics{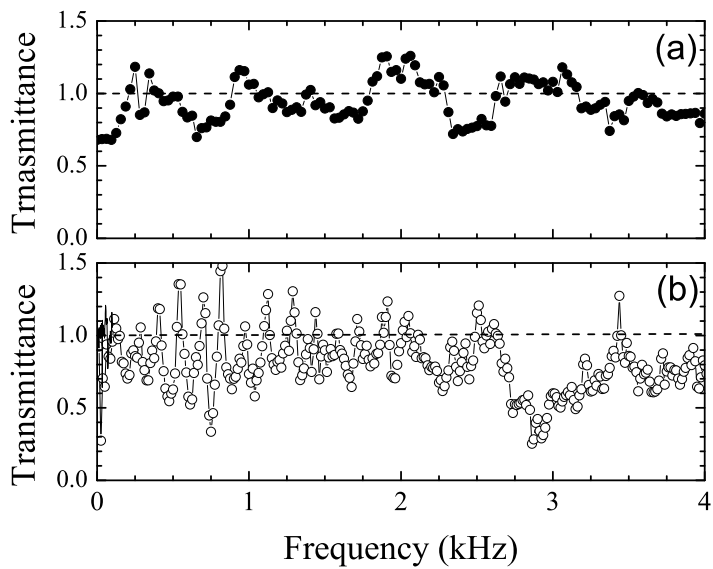}
\caption{\label{fig:MedidaPlancha}(a) Transmittance of a flat perforated metal plate 0.5 mm thick. The perforations are circular with 0.5mm diameter and are arranged in a hexagonal lattice.(b) Transmittance through a SC structure consisting of three rows of cylinders (8 cm diameter) fabricated with the perforated plates characterized above. The dashed horizontal lines are guides for the eye.}
\end{figure}
%
%
\begin{figure}
\includegraphics{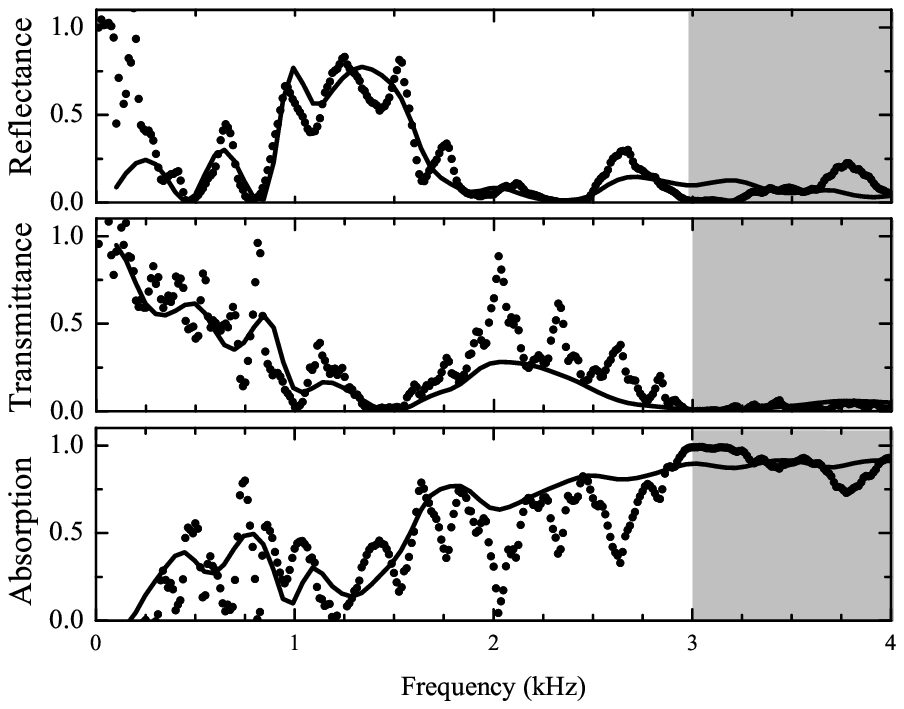}
\caption{\label{fig:GraficaG+NG}Reflectance, transmittance and absorption spectra (black dots) for a noise barrier made of cylinders with rubber crumb layer (8 cm diameter) and a rigid core (4 cm diameter) (sample 3 in table I). The corresponding simulations (solid lines) are obtained by the multiple scattering theory described in Sec. II.}
\end{figure}
%
%
\begin{figure}
\includegraphics{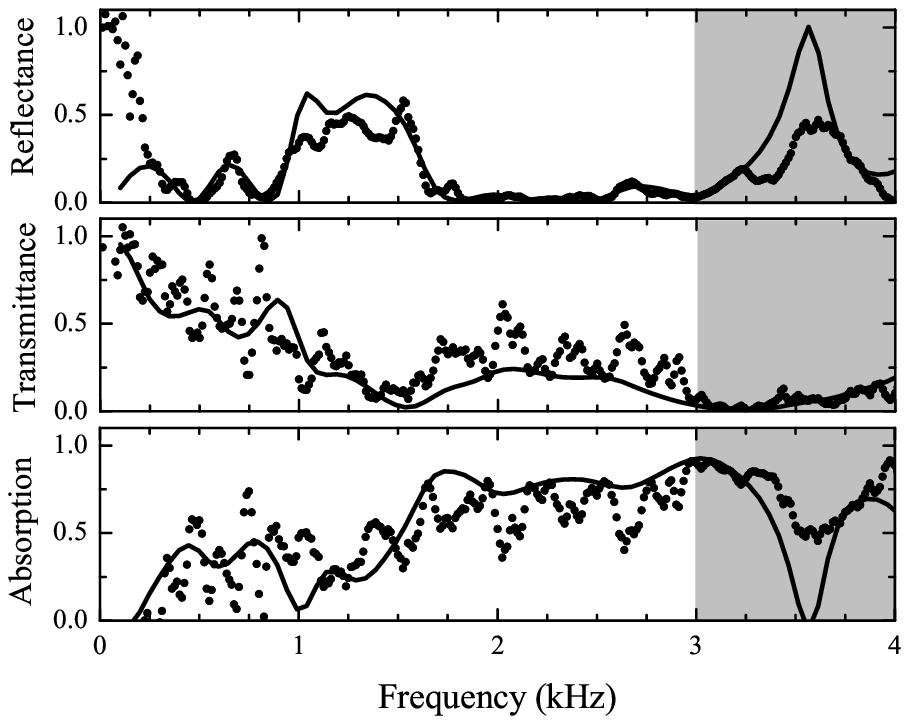}
\caption{Reflectance, transmittance and absorption spectra (black dots) for a noise barrier made of cylinders with rubber crumb layer (8 cm diameter) and a rigid core (2 cm diameter) (sample 4 in table I). The corresponding simulations (solid lines) are obtained by the multiple scattering theory described in Sec. II.}
\label{fig:GraficaG+NF}
\end{figure}
%
%
\begin{figure}
\includegraphics{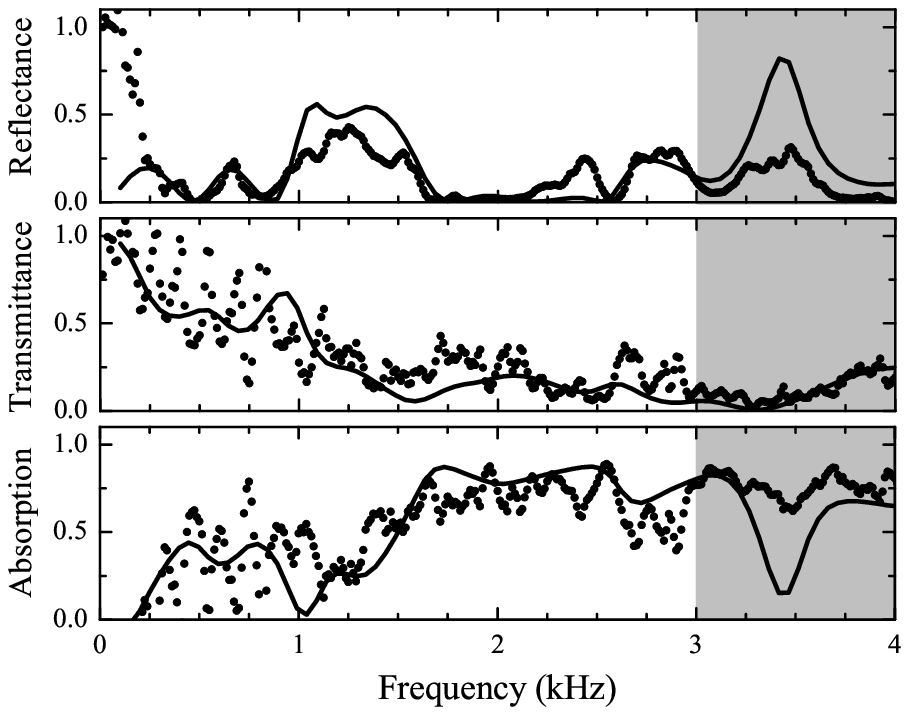}
\caption{\label{fig:GraficaG10}Reflectance, transmittance and absorption spectra (black dots) for a noise barrier made of cylinders with rubber crumb layer (8 cm diameter) (sample 5 in table I). Numerical simulations (continuous lines) are obtained by the multiple scattering theory described in Sec. II.}
\end{figure}
The rubber crumb cylinders are fabricated by inserting this recycled material inside hollow cylinders made with 0.5 mm thick steel plates perforated with holes of 0.5 mm of diameter and arranged in a hexagonal lattice with 1.3mm of lattice constant. This procedure represents a practical alternative to the habitual use of binders in order to obtain rigid structures of rubber crumb. Experimental verification that the described perforated metal plates are highly transparent to sound is shown in Fig. \ref{fig:MedidaPlancha}(a), which represents the transmission coefficient measured at 10cm from the rear surface by using the transmission set up explained previously. From this spectrum it can be concluded that the plate is almost acoustically transparent for normal incident sound. If sound impinges with angles different to the normal, the plate loses its transparency and, as a consequence, the cylinders made with them loose the perfect transparency for frequencies above the diffraction limit of the associate SC (about 2.8 kHz). This phenomenon is observed in the transmittance spectrum depicted in Fig. \ref{fig:MedidaPlancha}(b), which corresponds to a barrier consisting of three layers of hollow cylinders 8cm diameter made with the perforated metal plates. Note that this structure is transparent enough until 4.0kHz. Therefore, it will be assumed that the absorption effects observed after their filling in will be associated solely to the absorption properties of the rubber crumb.

Figures \ref{fig:GraficaG+NG}, \ref{fig:GraficaG+NF}, and \ref{fig:GraficaG10} represent the spectra for samples 3, 4 and 5, respectively. Note that, as in the case of SC barriers based on rigid cylinders, we have obtained an overall good agreement between measurements (black dots) and numerical simulations (continuous lines). In comparison with the case of samples 1 and 2, there is now an important absorption in the full range of frequencies because of the attenuation properties of rubber crumb. Let us remark that absorption increases for increasing frequency because of the behavior of $k_c$ shown in Fig. \ref{fig:KZgranza}(b). Also note that no unphysical results are observed in reflectance and transmittance spectra above the diffraction limit since the absorption mechanism avoid their appearance. However, the data above this limit are of no quality and must be obtained by other means.

The reflectance spectra in Figs. \ref{fig:GraficaG+NG}, \ref{fig:GraficaG+NF}, and \ref{fig:GraficaG10} show a relevant feature: the peaks at the band gap region have approximately the same frecuency width due to the fact that all the structure has the same filling ratio; i.e., the cylinders have the same external diameter. However, the peak is stronger for the case of sample 3, which is made of cylinders with the thicker core, and weaker for sample 4, which is made of cylinders with no core. 

Regarding transmittance and absorption spectra, it is difficult to extract clear conclusions by just looking at their corresponding plots in Figs. \ref{fig:GraficaG+NG}, \ref{fig:GraficaG+NF}. So, these magnitudes are discussed in the next section with the help of related parameters.

\subsection{Discussion}
\begin{figure}
\includegraphics{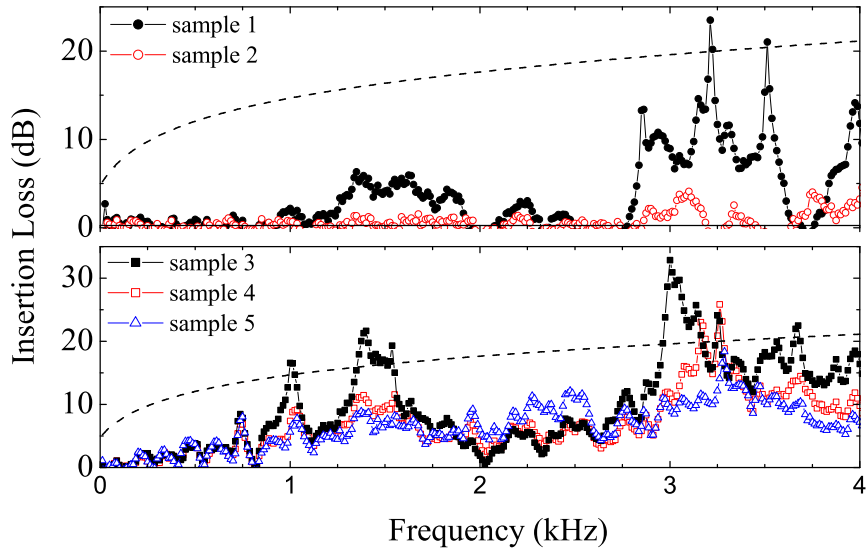}
\caption{\label{fig:ILoss1}(Color online) Insertion loss of samples described in Table I.
The dashed lines represent the IL predicted for a rigid wall having the same external dimensions as the samples.}
\end{figure}
\begin{figure}
\includegraphics{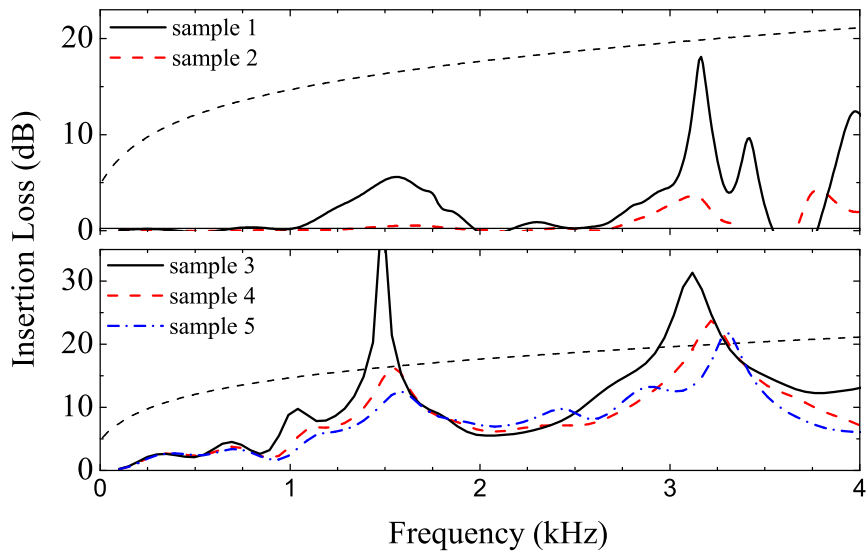}
\caption{\label{fig:ILossTheo}(Color online) Theoretical predictions of the Insertion loss for the samples described in Table I. 
The dashed lines represent the IL predicted for a rigid wall having the same external dimensions as the samples.}
\end{figure}
In order to compare the absorption quality of the samples analyzed we have introduced $A_P$, a parameter defined as
\begin{equation}
\label{eq:qua}
A_P\equiv\frac{1}{\Delta \omega}\int_{\Delta \omega}A(\omega)d\omega,
\end{equation}
where $A(\omega)$ is the energy absorption. $A_P$ acts as a quality parameter to measure the absorption power at the three different spectral regions of interest in a SC: below bandgap, at bangap and above bandgap. The borders of the bandgap are defined by the frequencies at which the reflectance peak is half of its maximum value. Note the $A_P=$1 is the maximum value achievable and would mean total absorption of the incident sound in the frequency range considered.  

The values of $A_P$ for the different samples are reported in Table \ref{tab:Quality}. 
First, let us remark that negative values mean that the energy conservation assumption is broken and, therefore, the resulting values for $A_P$ have no physical meaning and are not discussed. 
For the case of barriers made of rigid cylinders (samples 1 and 2) there is no any physical mechanism leading to sound absorption and, therefore, $A_P$ values should always be close to zero.   
Results for samples 3, 4 and 5 indicate that power absorption below the bandgap are practically the same and does not depend on the thickness of the porous covering. Inside the bandgap, the experiment show that $A_P$ increases with increasing shell thickness; it is maximum when the cylinders have no core. 
For frequencies above bandgap $A_P$ get larger values since attenuation by rubber crumb increases with frequency but, in this region, we have to recall that data obtained by Eq. \eqref{R_2micros} are not reliable and a more accurate method should be employed in this region. Therefore, for frequencies above bandgap, $A_P$ must be calculated by using data obtained by a different experimental set up.
Table \ref{tab:Quality} also shows that there is a general good agreement between experimental data and simulations based on a 2D multiple scattering algorithm. Let us remark that the disagreement observed at the bandgap region are mainly due to the overestimation by our modeling of the peak strength in reflectance spectra.

The insertion loss (IL) of the different samples either measured as well as calculated are depicted in Figs. \ref{fig:ILoss1} and \ref{fig:ILossTheo}, respectively, for the sake of their comparison with results reported by other groups\cite{San02,Gof03,UmnJASA05}. The black dashed lines in these figures represent the IL calculated for a rigid wall with the same dimensions than the SC barriers; i.e., 30 cm thick and 1 m height. This value has been obtained by using the procedure described in the ISO 9613\cite{ISO9613}. It is observed that the IL of samples containing a porous shell are stronger than that having only rigid cylinders. This behavior is produced by the absorption properties of the rubber crumb. Moreover, it is also observed that the shell containing the thicker core present an enhancement of the IL, which even overcome the IL of the corresponding rigid wall in certain frequency regions. This result is confirmed by the simulations and let us to conclude that a critical thickness that optimizes the IL in this type of structures may exist.

On one hand, the comparison with the attenuation properties of SC barriers based on rigid cylinders\cite{San02,Gof03} indicates that IL values comparable to ours could be obtained by increasing the number of layer employed in building the barriers. However, this IL enhancement will be focused into the band gap frequency region. Note that in our rubber crumb based structures the attenuation is obtained in a broad range of frequencies.
On the other hand, the comparison with the barriers using the same attenuation mechanisms \cite{UmnJASA05} than that employed here is difficult because of several reasons: (1) the differences in the geometrical parameters of the barriers studied, (2) the porous material employed are also different, and (3) the low frequency region in Ref. \onlinecite{UmnJASA05} is not explicitly analyzed. In any case, it could be said that the IL in our porous structure is stronger (in the low frequency region) than that reported in Ref. \onlinecite{UmnJASA05} since thicker layers of porous materials are better low frequency absorbers. 

The results shown in Figs. \ref{fig:ILoss1} and  \ref{fig:ILossTheo} are extremely interesting because they indicate that (at low frequencies) the IL dramatically increases when we add a layer of rubber crumb to the metallic cylinders building the barrier. For example, we observe that the IL for the barrier consisting of cylinders made of a metallic core (4cm diameter) and a rubber crumb layer with a thickness of 2 cm is about three times larger (at the bandgap frequencies) than that fabricated only with metallic cores. 
The IL enhancement observed at all frequencies is due to two main effects: the higher filling fraction of the SC barrier caused by the overlayer of rubber crumb and the dissipation produced by its porosity, which increases with increasing frequencies according to the properties of rubber crumb described in Fig. \ref{fig:KZgranza}.

The dashed lines in Figs. \ref{fig:ILoss1} and \ref{fig:ILossTheo} represent the IL of a rigid wall having the same external dimensions than SC samples. It has been calculated by using the expressions in Ref. \onlinecite{ISO9613}. It is observed that SC barriers based on only three rows of rigid cylinders show a very low attenuation efficiency in comparison with the rigid wall. The attenuation efficiency of SC barriers is strongly enhanced by using porous cylinders as building blocks; their qualities approach the one of a rigid wall in some frequency regions.
Let us remark that IL similar to that of the rigid wall could be obtained by using additional rows cylinders.

Finally, it is interesting to remark a recent work predicting the quenching of acoustic band gaps by flow noise generated by wind impinging the barriers based on sonic crystals\cite{ElnAPL09}. Regarding this point, we should point out that wind speeds needed to destroy the band gaps is pretty high (above 10m/sec) and depend on the barrier filling fraction. Since the prototype barriers tested were 25cm long in height\cite{ElnAPL09}, we expect that for barriers of about 3m height and having a filling ratio (i.e., the ratio between the volume occupied by cylinders and the total volume)  of 41.5\%, sound attenuation by reflectance is guarantee for wind speeds below 30m/sec, which is obtained by using the relation of the wind speed expected for actual structures ($v$) with that measured on prototypes ($v_p$): $v=v_p\sqrt{\ell_/\ell_p}$, where $\ell_p$ is the dimension of prototype and $\ell$ the real dimension\cite{MunBook02}. The reader is addressed to chapter 7 of Ref. \onlinecite{MunBook02} for the origin and details of this relationship.
%
\section{Summary}
%
Reflectance and transmittance spectra have been measured and simulations are performed for noise barriers based on SCs made of three different types of cylindrical scatterers: rigid, porous and porous shells with a rigid core. Our model, which is based on the multiple scattering theory, accurately describes the absorption properties of the structures experimentally studied and also gives support to the measurement method employed. Our results indicate that three rows of cylinders are enough to get well defined bandgaps in the transmittance and reflectance spectra. This is a relevant finding in order to make the building of these type of noise barriers affordable. 

We have also shown that the effect of using rubber crumb like porous absorptive media results in an enhancement of the insertion loss of these barriers  in comparison with those based on just rigid cylinders. The IL enhancement have been demonstrated in the low frequency regime (below 3 kHz) analyzed here.  
Moreover, it has been demonstrated that the inner structure of the porous cylinders (i.e., the possible existence of a rigid core) is a mechanism that can be used to optimize the IL of SC barriers at different frequency regions. 
Therefore, an optimization procedure can be developed in order to design SC barriers based on building units efficiently adapted to attenuate sound for different noisy environments. In this regards, traffic noise is a paramount example where that optimization procedure should be applied first to test the ability of rubber crumb in shielding broadband noise with low frequencies. Further work will be developed in our group to achieve this goal.
%
\begin{acknowledgments}
%
This work has been partially supported by the Spanish MICINN under grants No. TEC2007-67239 and No. CSD2008-66 (CONSOLIDER Program). DT acknowledges a research contract provided by the program Campus de Excelencia 2010 UPV. The authors acknowledge F. Payri and A. Broatch for providing the technical facilities that made possible some of the measurements. 
\end{acknowledgments}





\begin{table}
\caption{\label{tab:Samples} Dimensions of the cylindrical units used in building the barrier samples studied here. $d_b$ is the external diameter of the cylindrical porous shell and $d_a$ is the diameter of its inner core, which is a rigid cylinder. The column $R_b-R_a$ gives the thicknesses of the porous shells. The last column reports the filling fraction $ff$ of the square lattice; $ff=\pi R_b^2/a^2$, where $a=$11cm is the lattice constant.}
\begin{ruledtabular}
\begin{tabular}{ccccc}
&{$d_a$({\rm cm})} & $d_b$(cm) &{$R_b-R_a$} & ff\\
\hline
sample 1 & 4 & 4 & 0 & 0.10\\
sample 2 & 2 & 2 & 0 & 0.03\\
sample 3 & 4 & 8 & 2 & 0.41\\
sample 4 & 2 & 8 & 3 & 0.41\\
sample 5 & 0 & 8 & 4 & 0.41\\
\end{tabular}
\end{ruledtabular}
\end{table}

\begin{table}
\caption{\label{tab:Quality} The absorption power $A_P$ [see Eq. \eqref{eq:qua}] for the barriers analyzed. This quality factor is calculated separately in the three frequency regions where the associated sonic crystals can be divided. The sample description is given in table \ref{tab:Samples} and Section \ref{sec:ExpCharac}. Negative values have no physical meaning (see text)}
\begin{ruledtabular}
\begin{tabular}{p{2cm}cccccc}
& \multicolumn{2}{c}{Below Gap}& \multicolumn{2}{c}{Band gap}&
\multicolumn{2}{c}{Above Gap}\\
&Theo.&Exp.&Theo.&Exp.&Theo.&Exp.\\
\hline
sample 1&$0$   &$0.04$&$-0.03$&$0.14$& -     & -  \\	
sample 2&$0$   &$0.01$&$-0.08$&$0.05$& -     & -  \\	
sample 3&$0.37$&$0.35$&$0.24$&$0.27$&$0.78$&$0.77$\\	
sample 4&$0.37$&$0.36$&$0.29$&$0.39$&$0.64$&$0.68$\\	
sample 5&$0.36$&$0.36$&$0.28$&$0.43$&$0.68$&$0.72$\\	
\end{tabular}
\end{ruledtabular}
\end{table}

\end{document}